\documentclass[aps,superscriptaddress,amsmath,amssymb]{revtex4-1}

\usepackage{graphicx}
\usepackage{bm}

\usepackage[T1]{fontenc}
\usepackage[utf8]{inputenc}
\usepackage{amsmath}
\usepackage{float}
\usepackage{color} 

\draft

\begin{document}

\title[Semimetal to semiconductor transition in $\text{Bi}/\text{TiO}_{2}$ core/shell nanowires]{Semimetal to semiconductor transition in $\text{Bi}/\text{TiO}_{2}$ core/shell nanowires}

\author{M. Kockert}
\email{kockert@physik.hu-berlin.de}
\affiliation{Novel Materials Group, Humboldt-Universit\"{a}t zu Berlin, 10099 Berlin, Germany}
\author{R. Mitdank}
\affiliation{Novel Materials Group, Humboldt-Universit\"{a}t zu Berlin, 10099 Berlin, Germany}
\author{H. Moon}
\affiliation{Department of Material Science and Engineering, Yonsei University, 03722 Seoul, Republic of Korea}
\author{J. Kim}
\affiliation{Division of Nanotechnology, DGIST, 42988 Daegu, Republic of Korea}
\author{A. Mogilatenko}
\affiliation{Ferdinand-Braun-Institut, Leibniz-Institut f\"{u}r H\"{o}chstfrequenztechnik, 12489 Berlin, Germany}
\author{S. H. Moosavi}
\affiliation{Laboratory for Design of Microsystems, University of Freiburg - IMTEK, 79110 Freiburg, Germany}
\author{M. Kroener}
\affiliation{Laboratory for Design of Microsystems, University of Freiburg - IMTEK, 79110 Freiburg, Germany}
\author{P. Woias}
\affiliation{Laboratory for Design of Microsystems, University of Freiburg - IMTEK, 79110 Freiburg, Germany}
\author{W. Lee}
\affiliation{Department of Material Science and Engineering, Yonsei University, 03722 Seoul, Republic of Korea}
\author{S. F. Fischer}
\email{saskia.fischer@physik.hu-berlin.de}
\affiliation{Novel Materials Group, Humboldt-Universit\"{a}t zu Berlin, 10099 Berlin, Germany}


\begin{abstract}

We demonstrate the full thermoelectric and structural characterization of individual bismuth-based (Bi-based) core/shell nanowires. The influence of strain on the temperature dependence of the electrical conductivity, the absolute Seebeck coefficient and the thermal conductivity of bismuth/titanium dioxide ($\text{Bi}/\text{TiO}_{2}$) nanowires with different diameters is investigated and compared to bismuth (Bi) and bismuth/tellurium (Bi/Te) nanowires and bismuth bulk. Scattering at surfaces, crystal defects and interfaces between the core and the shell reduces the electrical conductivity to less than $5\,\%$ and the thermal conductivity to less than $25\,\%$ to $50\,\%$ of the bulk value at room temperature. On behalf of a compressive strain, $\text{Bi}/\text{TiO}_{2}$ core/shell nanowires show a decreasing electrical conductivity with decreasing temperature opposed to that of Bi and Bi/Te nanowires. We find that the compressive strain induced by the $\text{TiO}_{2}$ shell can lead to a band opening of bismuth increasing the absolute Seebeck coefficient by $10\,\%$ to $30\,\%$ compared to bulk at room temperature. In the semiconducting state, the activation energy is determined to $\left|41.3\pm0.2\right|\,\text{meV}$. We show that if the strain exceeds the elastic limit the semimetallic state is recovered due to the lattice relaxation.

\end{abstract}

\maketitle

\section{\label{sec:introduction}Introduction}

Bismuth (Bi) has been under investigations for a long time \cite{Gallo} due to its unique properties, e.g. its anisotropic transport properties, long charge carrier mean free path (up to a few hundred micrometers at $4\,\text{K}$), large Fermi wavelength ($70\,\text{nm}$) and semimetal band structure \cite{Gallo,CroninLin,DuggalRup,Liu,ZhangSun,HeremansThrush,LeeHam}. However, Bi bulk has a low thermoelectric performance \cite{Gallo,Dresselhaus}, given by the figure of merit $ZT=\frac{\sigma S^{2} T}{\lambda}$, where $\sigma$ is the electrical conductivity, $S$ is the absolute Seebeck coefficient and $\lambda$ is the thermal conductivity at a certain bath temperature $T$. Dresselhaus \textit{et al.} \cite{Dresselhaus} predicted theoretically an improvement of the thermoelectric efficiency due to quantum size effects for 1D quantum-wire-structures made of Bi. However, the practical implementation of Bi nanowires with such small diameters into applications can be challenging. For this reason, Bi-based core/shell nanowires have raised attention in recent years because of their increased thermoelectric performance for relatively large diameters ($d>300\,\text{nm}$) \cite{KangRoh,KimKim,KimOh}. Tellurium (Te) coated as shell on the Bi nanowire core can cause a semimetal to semiconductor transition due to the lattice mismatch of the Bi core and the Te shell \cite{KimKim,KimOh}. An increased Seebeck coefficient and a reduced thermal conductivity can be the result of such a heterostructure. However, the combined full thermoelectric characterization of individual Bi-based core/shell nanowires, in which all transport parameters ($\sigma$, $S$ and $\lambda$) are determined on one and the same nanowire, remains an open issue. Here, we demonstrate the full thermoelectric transport and structural characterization of individual core/shell nanowires \cite{Kojda1,Kojda2,KockertAg} and investigate titanium dioxide ($\text{TiO}_{2}$) as insulating shell material for Bi-based nanowires and compare this to Bi and Bi/Te nanowires. We detect the transition from the semiconducting to semimetallic state and vice versa and discuss the influence of the shell material on the thermoelectric properties of individual Bi-based nanowires.

\section{\label{sec:exp-details}Experimental Details}

\begin{table}[htbp]
	\centering
		\begin{tabular}{|c|c|c|c|}
			\hline 
			Sample & Diameter $d\,(\text{nm})$ & Length $l\,(\mu\text{m})$ & Shell thickness $t\,(\text{nm})$\\
			\hline 
			Bi 1 & $170 \pm 5$ & $15.0 \pm 0.6$ & $-$\\
			Bi 2 & $210 \pm 5$ & $15.4 \pm 0.7$ & $-$\\
			Bi 3 & $550 \pm 10$ & $11.3 \pm 0.7$ & $-$\\
			\hline 
			$\text{Bi}/\text{TiO}_{2}$ 1 & $155 \pm 5$ & $17.1 \pm 1.8$ & $30$\\
			$\text{Bi}/\text{TiO}_{2}$ 2 & $470 \pm 10$ & $15.4 \pm 1.1$ & $30$\\
			$\text{Bi}/\text{TiO}_{2}$ 3 & $590 \pm 10$ & $15.5 \pm 1.3$ & $30$\\
			\hline
		\end{tabular}
		\caption{\textbf{Geometry parameters.} Overview of entire diameter $d$, length $l$ and shell thickness $t$ of bismuth (Bi) and bismuth/titanium oxide ($\text{Bi}/\text{TiO}_{2}$) nanowires, respectively. Bi nanowires have a native oxide layer of $5\,\text{nm}$ to $10\,\text{nm}$ \cite{RohHippalgaonkar,ShimHam}. $\text{Bi}/\text{TiO}_{2}$ nanowires are coated with a uniform $\text{TiO}_{2}$ shell with a thickness of $30\,\text{nm}$. The geometry parameters have been determined by scanning and transmission electron microscopy.}
		\label{tab:table}
\end{table}

The bismuth-based core/shell nanowires consist of a bismuth (Bi) core and a tellurium (Te) or titanium dioxide ($\text{TiO}_{2}$) shell, respectively. The single crystalline Bi core was prepared by means of the on-film formation of nanowires (OFFON) method as reported in Ref. \cite{KangRoh,KimKim,KimOh,ShimHam}. Bi thin films were deposited by radio frequency sputtering on $\text{SiO}_{2}/\text{Si}$ substrates. The sputtering system was evacuated to $10^{-7}$ Torr before the deposition \cite{KimOh}. The vacuum was maintained during the sputtering under a 2-mTorr Ar environment at a temperature of $300\,\text{K}$ \cite{KimOh}. After the deposition process, a heat treatment of the Bi thin films was conducted for several hours at $523\,\text{K}$ in a vacuum of $10^{-7}$ Torr \cite{KimOh}. A compressive thermal stress in the Bi film is induced due to the mismatch of the thermal expansion coefficients of the Bi thin film and the $\text{SiO}_{2}/\text{Si}$ substrate \cite{ShimHam}. This leads to the growth of the Bi nanowires. Radio frequency sputtering was then used to deposit the Te shell onto the Bi core. Atomic layer deposition was used to deposit $\text{TiO}_{2}$ as a shell material. All processes were performed in a high vacuum environment to prevent the formation of an oxidation layer between the core and the shell material.

A thermoelectric nanowire characterization platform (TNCP) \cite{Moosavi,Wang} was used to perform a full analysis of the transport properties of individual Bi-based core/shell nanowires. An electron transparent gap (width: $10\,\mu\text{m}$ to $20\,\mu\text{m}$) serves as thermal insulation of the suspended nanowire and enables an investigation of the specimen by scanning (SEM) and transmission electron microscopy (TEM). The measurement zone is situated in the middle of the platform. A sketch of this area is given in Fig. \ref{fig:REM_TEM_EDX}a. The TNCP consists of an insulating silicon dioxide surface. On top of that surface, $200\,\text{nm}$ platinum electrodes were prepared by radio frequency sputtering. Individual Bi-based nanowires were picked up from the growth substrate and placed on the measurement zone of the TNCP by a micromanipulation system with a thin tungsten tip. Electron beam-induced deposition (EBID) of platinum-based or tungsten-based precursor contacts was conducted in order to prepare a mechanical and electrical connection between the nanowire and the TNCP. For Bi and Bi/Te nanowires, the EBID contacts were applied directly on the shell material, see Fig. \ref{fig:REM_TEM_EDX}b. For $\text{Bi}/\text{TiO}_{2}$ core/shell nanowires, the shell was removed selectively by means of focused ion beam milling before the deposition of the EBID contacts, see Fig. \ref{fig:REM_TEM_EDX}c. 

A four-terminal configuration of the platinum electrodes depicted as $\text{E}_{\text{c}}$, $\text{E}_{\text{h}}$, $\text{T}_{\text{c}}$ and $\text{T}_{\text{h}}$ in Fig. \ref{fig:REM_TEM_EDX}a was used to measure the resistance $R$ of the Bi-based nanowires. 
The electrical conductivity $\sigma$ of Bi-based nanowires can be determined under the assumption that the cross-sectional area of the nanowires is circular by  

\begin{equation}
	\sigma=\frac{4l}{R\pi d_{e}^{2}},
\end{equation}

where $R$ is the four-terminal resistance of the nanowire, $l$ is the length and $d_{e}$ is the effective diameter. The effective diameter is smaller than the entire diameter $d$ due to the electrical insulating native oxide layer of the Bi nanowires and the electrical insulating $\text{TiO}_{2}$ shell of the $\text{Bi}/\text{TiO}_{2}$ nanowires. The uncertainty of the electrical conductivity $\sigma$ mainly comes from the determination of the geometry parameters. The diameter was measured by scanning electron microscopy (SEM) at several points along each nanowire. The uncertainty of the diameter results from the resolution limitation of the SEM investigations and from the diameter variation of the nanowires and is between $5\,\text{nm}$ and $20\,\text{nm}$. The length $l$ was also measured by SEM. The uncertainty of $l$ mainly comes from the size of contact area that is defined by the electron beam-induced deposition contacts and varies between $0.4\,\mu\text{m}$ and $1.8\,\mu\text{m}$. The four-terminal resistance $R$ was determined by linear fits of corresponding $I$-$V$ curves and its relative uncertainty is less than $1\,\%$. All geometry parameters are given in Tab. \ref{tab:table}. The resistance of the Bi-based nanowires as a function of the bath temperature is given in the Supplementary Information.

The temperature-dependent thermovoltage $U_{\text{S-Bi-based,Pt}}$ of individual Bi-based nanowires relative to $200$\,nm thick platinum conduction lines was measured between bath temperatures of $10\,\text{K}$ and $350\,\text{K}$. The temperature difference $\delta T$ between the hot and the cold side of the nanowires was calculated by the change of four-terminal-resistance thermometers due to the variation of the power of the micro heater on the TNCP by increasing the applied heating current $I_{\text{H}}$ from zero to $-I_{\text{H,max}}$ and from zero to $+I_{\text{H,max}}$ stepwise in equidistant steps. The slope of the function $U_{\text{S-Bi-based,Pt}} (\delta T)$ gives the relative Seebeck coefficient $S_{\text{Bi-based,Pt}}$ of the Bi-based nanowires with respect to the platinum conduction lines

\begin{equation}
	S_{\text{Bi-based,Pt}}=S_{\text{Bi-based}}-S_{\text{Pt}}=-\frac{\text{d} U_{\text{S-Bi-based,Pt}}}{\text{d} \delta T}.
	\label{eq:Seebeckrelativ}
\end{equation}

The absolute Seebeck coefficient of a Bi-based nanowire is given by

\begin{equation}
	S=S_{\text{Bi-based}}=S_{\text{Bi-based,Pt}}+S_{\text{Pt}},
	\label{eq:Seebeckabsolut}
\end{equation}

where $S_{\text{Pt}}$ is the absolute Seebeck coefficient of the platinum reference material. For bath temperatures between $T=10\,\text{K}$ and $T=300\,\text{K}$, $S_{\text{Pt}}$ was determined in a separate experiment \cite{KockertPt} by measuring a bulk gold wire with known absolute Seebeck coefficient relative to a thin platinum conduction line. For bath temperatures above $T=300\,\text{K}$, $S_{\text{Pt}}$ was taken from bulk platinum \cite{HuebenerPt1,HuebenerPt2}. This is reasonable because the difference between $S_{\text{bulk}}$ and $S_{\text{film}}$ is within the measurement limit. The uncertainty of the thermovoltage is given by the confidence interval of the thermovoltage which was measured ten times at each step of the applied heating current and then arithmetically averaged. The uncertainty of the relative Seebeck coefficient is determined by the modulus of the largest deviation of the mean value of different fit lines due to applied heating current $I_{\text{H}}$ that was varied from zero to $-I_{\text{H,max}}$ and from zero to $+I_{\text{H,max}}$. The uncertainty of the absolute Seebeck coefficient was determined by error propagation. 

The temperature-dependent thermal conductivity $\lambda$ of individual Bi-based nanowires was determined by the increase of the resistance of the nanowires due to self-heating \cite{VoelkleinReith}. A current was applied at the outer electrodes $\text{E}_{\text{c}}$ and $\text{E}_{\text{h}}$, see Fig. \ref{fig:REM_TEM_EDX}a, and gradually increased. The thermal conductivity $\lambda$ is given by 

\begin{equation}
	\lambda=\frac{1}{12}\frac{\alpha l R}{A}\frac{P}{\Delta R(P)}.
\end{equation}

$\alpha$ is the temperature coefficient of the resistance of the nanowire, $R$ is the four-terminal resistance, $l$ is the length, $P$ is the resulting power in the nanowire based on the voltage drop due to the applied current, $A$ is the cross-sectional area of the nanowire. The uncertainty of the thermal conductivity due to the different shell materials depends on the thermal conductivity of the shell material and the cross-sectional area occupied by the shell and will be discussed later.

In the experiments, the four-terminal resistance of the nanowires was measured by a Keithley 6221 AC and DC Current Source and a Keithley 2182A Nanovoltmeter. For Seebeck measurements, the micro heater power was controlled by a Keithley SourceMeter 2401. The thermometer resistances were determined by four-terminal measurements performed by Keithley 6221 and 2182A devices. The thermovoltage was measured by a Keithley 2182A Nanovoltmeter. The measurement configurations were changed by a Keithley 7001 switch matrix system. The transport experiments of the Bi/Te and $\text{Bi}/\text{TiO}_{2}$ core/shell nanowires were performed in a flow cryostat in helium atmosphere at ambient pressure for the electrical and Seebeck measurements and in vacuum for the thermal conductivity measurements, respectively. All transport experiments of the Bi nanowires were performed in a closed cycle cryocooler in vacuum. Scanning (SEM) and transmission electron microscopy (TEM) as well as energy-dispersive X-ray spectroscopy (EDX) were performed to investigate the structure and chemical composition of the Bi-based core/shell nanowires. 

\section{\label{sec:results_discussion}Results and discussion}

\subsection{Structural properties}

\begin{figure}[htbp]
\includegraphics[width=0.85\textwidth]{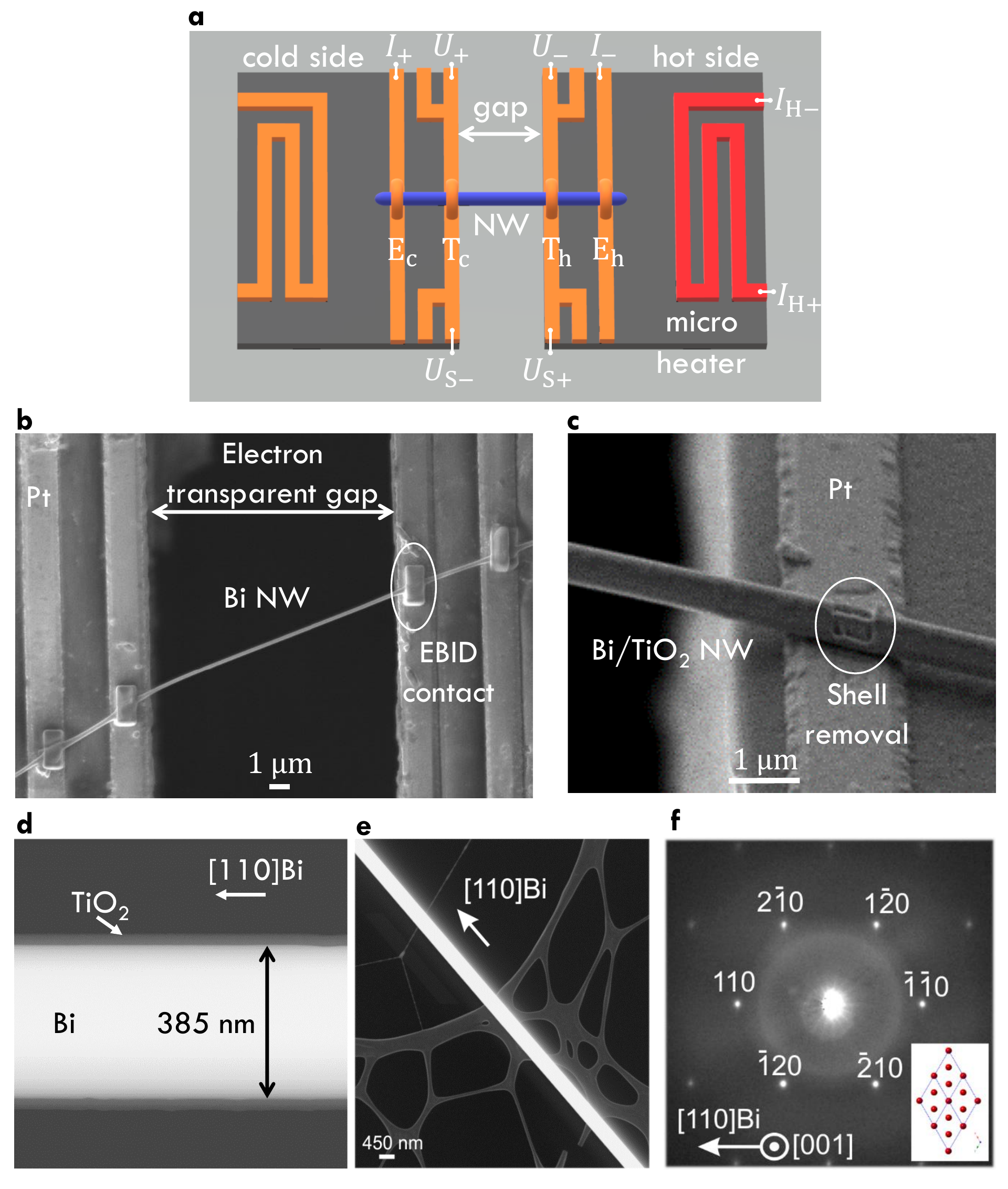}
\caption{\textbf{Structural properties of the Bi-based core/shell nanowires.} \textbf{a}, Sketch of the measurement area of the platform. An electron transparent gap divides the measurement area into two sides. The determination of the four-terminal resistance of a nanowire (NW) (blue) can be performed by applying a current $I$ at the outer platinum conduction lines $\text{E}_{\text{c}}$ and $\text{E}_{\text{h}}$ (orange) and measuring the voltage $U$ at inner conduction lines. The thermovoltage $U_\text{S}$ of a nanowire relative to the platinum conduction lines can be measured by applying a heating current $I_\text{H}$ at the micro heater (red). This creates a temperature difference along the nanowire that can be calculated by four-terminal resistance thermometers $\text{T}_{\text{c}}$ and $\text{T}_{\text{h}}$ (orange) for the cold and hot side, respectively. \textbf{b}, Scanning electron microscopy image of a Bi nanowire (Bi 1) placed on the thermoelectric nanowire characterization platform. Electrical and mechanical connection between the nanowire and the measurement platform was prepared by electron beam-induced deposition (EBID). \textbf{c}, Scanning electron microscopy image of a $\text{Bi}/\text{TiO}_{2}$ nanowire after a selected area shell removal in order to prepare EBID contacts directly at the Bi core. \textbf{d}, Scanning transmission electron microscopy image of a $\text{Bi}/\text{TiO}_{2}$ showing the uniform shell thickness. \textbf{e}, Scanning transmission electron microscopy image of a $\text{Bi}/\text{TiO}_{2}$ nanowire placed on a carbon film showing the nanowire growth direction. \textbf{f}, Selected area electron diffraction confirms the single crystalline crystal structure of the Bi core of a $\text{Bi}/\text{TiO}_{2}$ nanowire. }
\label{fig:REM_TEM_EDX}
\end{figure}

Fig. \ref{fig:REM_TEM_EDX}b shows a Bi nanowire bridging the platform gap and attached with EBID contacts to four platinum conduction lines. Fig. \ref{fig:REM_TEM_EDX}c shows a SEM image of a $\text{Bi}/\text{TiO}_{2}$ core/shell nanowire before the contact preparation. The shell was removed selectively by focused ion beam milling because of the electrical insulating behavior of $\text{TiO}_{2}$. Fig. \ref{fig:REM_TEM_EDX}d shows an image of a $\text{Bi}/\text{TiO}_{2}$ nanowire with a uniform shell prepared by atomic layer deposition. A scanning transmission electron microscopy image (Fig. \ref{fig:REM_TEM_EDX}e) of a $\text{Bi}/\text{TiO}_{2}$ nanowire placed on a carbon film exhibits the growth direction. Fig. \ref{fig:REM_TEM_EDX}f depicts the selected area electron diffraction pattern of the $\text{Bi}/\text{TiO}_{2}$ nanowire proving its single crystallinity. Indexing the electron diffraction spots confirms the rhombohedral crystal structure of the Bi core (see the structural model in the inset of Fig. \ref{fig:REM_TEM_EDX}f). The geometry parameters of the Bi and $\text{Bi}/\text{TiO}_{2}$ nanowires are given in Tab  \ref{tab:table}. For comparison, our measurement data for Bi/Te nanowires are given in the Supplementary Information  

\subsection{Electrical characterization}

The temperature-dependent electrical conductivity $\sigma$ of Bi and $\text{Bi}/\text{TiO}_{2}$ nanowires is shown Fig. \ref{fig:Bi_Based_Elektrisch_Seebeck}a. Moreover, $\sigma_{\text{bulk}}$ (perpendicular to the trigonal axis) from Ref. \cite{Gallo} is added to the diagrams. The electrical conductivity of all nanowires is reduced compared to the bulk material. $\text{Bi}/\text{TiO}_{2}$ nanowires exhibit a clearly reduced electrical conductivity compared to the Bi nanowires and a semiconducting temperature dependence of the electrical conductivity.

The temperature dependence of the electrical conductivity of the bulk Bi semimetal can be explained by the competing influence of carrier concentration and mobility \cite{Gallo,ZhangSun,HeremansThrush,HuberNikolaeva,NikolaevaHuber,KimLee}. Bi bulk has a small carrier concentration varying between $2.7\cdot10^{17}\,\text{cm}^{3}-3.0\cdot10^{18}\,\text{cm}^{3}$ at temperatures between $2\,\text{K}-300\,\text{K}$ \cite{Gallo,ZhangSun,HeremansThrush,KimLee}. The change of mobility $\mu$ of Bi bulk exceeds the change of the carrier concentration by more than one order of magnitude in the temperature range from $77\,\text{K}-300\,\text{K}$ \cite{Gallo,ZhangSun,HeremansThrush}. 

All Bi nanowires show semimetallic behavior and have a reduced electrical conductivity compared to that of bulk. This can be attributed to enhanced surface scattering due to a higher surface-area-to-volume ratio in nanowires. Hence, $\sigma$ of Bi 1 ($170\,\text{nm}$) is reduced compared to that of Bi 2 ($210\,\text{nm}$). A smaller diameter leads to increased surface scattering, which will reduce the electrical conductivity, because the mean free path of the charge carriers becomes restricted by the nanowire diameter. This is consistent with Ref. \cite{KimLee,KimShim,NikolaevaHuber}. 

However, $\sigma$ of Bi 3 ($550\,\text{nm}$) shows a maximum at around $240\,\text{K}$ as was observed in Ref. \cite{ShimHam,WangZhang}. A detailed figure of the resistance is given in the Supplementary Information. This temperature dependence of Bi 3 ($550\,\text{nm}$) may be attributed to a change of the dominant scattering mechanism from surface scattering (Bi 1 ($170\,\text{nm}$) and Bi 2 ($210\,\text{nm}$)) to grain boundary scattering (Bi 3 ($550\,\text{nm}$)) below a certain temperature. Grain boundary scattering may arise in nanowires with larger diameters because of twin boundaries. From the temperature dependence of the electrical conductivity the modulus of the thermal activation energy \cite{Seto} of Bi 3 ($550\,\text{nm}$) is determined to $\left|2.0 \pm 0.1\right|\,\text{meV}$ in the temperature range from $100\,\text{K}$ to $200\,\text{K}$. 

According to Matthiessen's rule, the total scattering rate, which is given by $\tau_\text{tot}^{-1}=\tau_\text{bulk}^{-1}+\tau_\text{sc}^{-1}+\tau_\text{gb}^{-1}$, leads to a reduction of the charge carrier mean free path. $\tau_\text{bulk}^{-1}$ characterizes the inverse lifetime in the bulk material and its temperature dependence can be described by the Bloch-Gr\"{u}neisen relation. $\tau_\text{sc}^{-1}$ is the surface scattering rate which depends on the nanowire diameter. $\tau_\text{gb}^{-1}$ is the grain boundary scattering rate which depends on the thermal activation of charge carriers. 

For $\text{Bi}/\text{TiO}_{2}$ core/shell nanowires, an electrical conduction in the shell material can be neglected due to the electrical insulating $\text{TiO}_{2}$. All $\text{Bi}/\text{TiO}_{2}$ nanowires show a decreasing electrical conductivity with decreasing bath temperature. 
$\text{Bi}/\text{TiO}_{2}$ core/shell nanowires show an increasing electrical conductivity with decreasing diameter and exhibit a diameter-dependent transition from the semiconducting to the semimetallic state.

\begin{figure}[htbp]
\includegraphics[width=0.95\textwidth]{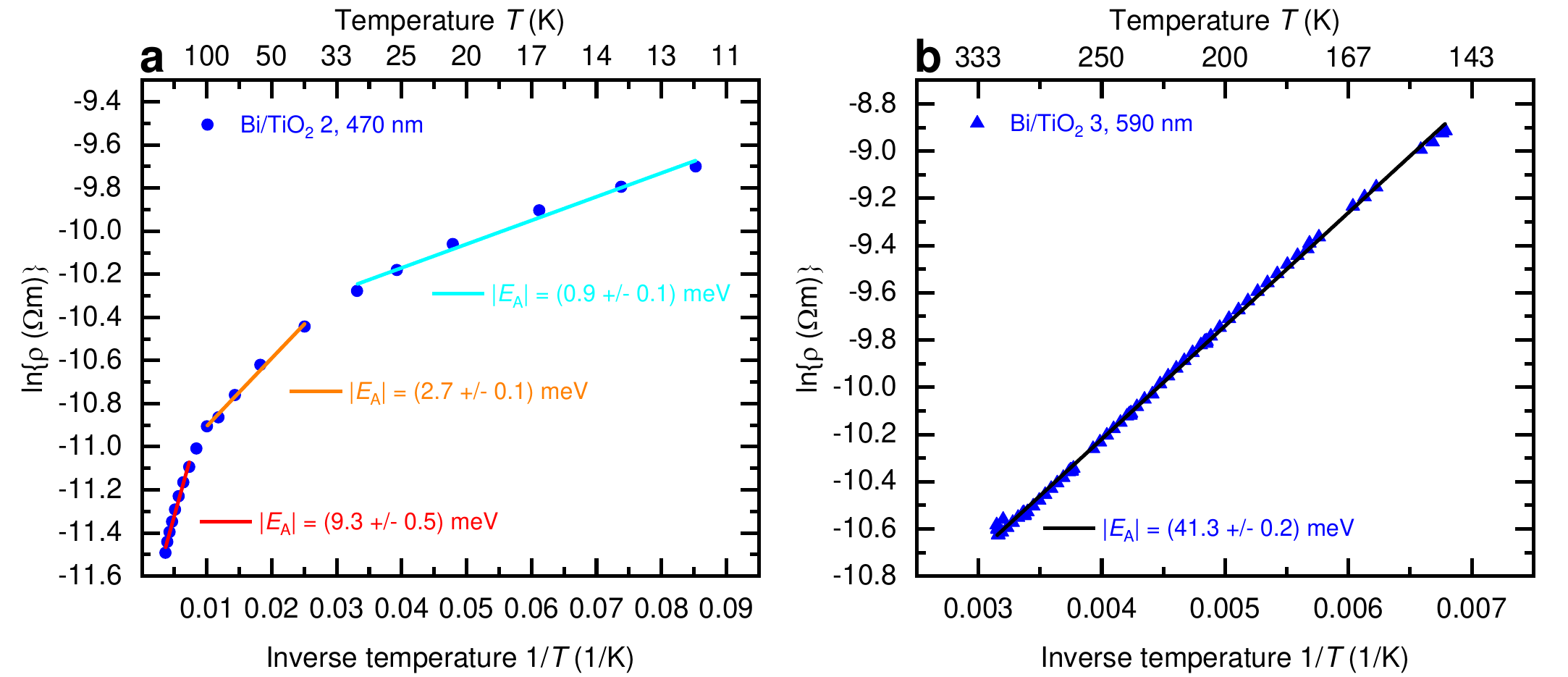}
\caption{\textbf{Activation energy of $\text{Bi}/\text{TiO}_{2}$ core/shell nanowires.} \textbf{a}, Natural logarithm of resistivity of $\text{Bi}/\text{TiO}_{2}$ 2 ($470\,\text{nm}$) as a function of the inverse bath temperature $T^{-1}$. The modulus of the activation energy $\left|E_\text{A}\right|$ can be determined in three distinct temperature ranges from $10\,\text{K}$ to $30\,\text{K}$, from $40\,\text{K}$ to $120\,\text{K}$ and from $140\,\text{K}$ to $270\,\text{K}$. Increasing the bath temperature also increases the activation energy. $\text{Bi}/\text{TiO}_{2}$ 2 ($470\,\text{nm}$) shows a semimetallic-like behavior. \textbf{b}, Natural logarithm of resistivity of $\text{Bi}/\text{TiO}_{2}$ 3 as a function of the inverse bath temperature $T^{-1}$. The activation energy $\left|E_\text{A}\right|$ can only be determined in the temperature range from $140\,\text{K}$ to $310\,\text{K}$. $\text{Bi}/\text{TiO}_{2}$ 3 ($590\,\text{nm}$) shows a semiconducting behavior.}
\label{fig:Aktivierungsenergie}
\end{figure}

The effect of the elastic strain of the $\text{TiO}_{2}$ shell can be observed for $\text{Bi}/\text{TiO}_{2}$ 3 ($590\,\text{nm}$) in Fig. \ref{fig:Bi_Based_Elektrisch_Seebeck}a. 
The electrical conductivity of $\text{Bi}/\text{TiO}_{2}$ 3 ($590\,\text{nm}$) is 25 times smaller than that of the bulk material and nearly 15 times smaller compared to the Bi nanowires at room temperature. The strain effect leads to an opening of a band gap. This is illustrated in Fig. \ref{fig:Aktivierungsenergie} which shows the natural logarithm of the resistivity of the $\text{Bi}/\text{TiO}_{2}$ nanowires as a function of the inverse bath temperature. The Arrhenius equation

\begin{equation}
	\rho=A\exp{\left(-\frac{E_{\text{A}}}{k_{\text{B}}T}\right)}
	\label{eq:Arrhenius}
\end{equation}

can be used to determine the activation energy $E_\text{A}$. $\rho$ is the resistivity, $A$ is a constant, $k_{\text{B}}$ is the Boltzmann constant and $T$ is the bath temperature. An activation energy of $\left|41.3\pm0.2\right|\,\text{meV}$ in the temperature range from $140\,\text{K}$ to $310\,\text{K}$ was determined for $\text{Bi}/\text{TiO}_{2}$ 3 ($590\,\text{nm}$). For $\text{Bi}/\text{TiO}_{2}$ 2 ($470\,\text{nm}$), a reduced activation energy of $\left|9.3\pm0.5\right|\,\text{meV}$ compared to  $\text{Bi}/\text{TiO}_{2}$ 3 ($590\,\text{nm}$) was determined.

When the diameter is further reduced down to the region of $\text{Bi}/\text{TiO}_{2}$ 1 ($155\,\text{nm}$), then the influence of the shell on the core exceeds the elastic limit and as a result the Bi core relaxes spontaneously and exhibits a semimetallic behavior of the electrical conductivity.

The relaxation process alters the electrical conductivity significantly, see Fig. \ref{fig:Bi_Based_vorher_nachher_quer}a. An irreversible increase of $\sigma$ of $\text{Bi}/\text{TiO}_{2}$ 3 ($590\,\text{nm}$) occurred during the measurement and the temperature dependence of $\sigma$ changed from semiconducting to semimetallic. A similar irreversible increase was observed for Bi/Te 3 ($490\,\text{nm}$). We infer that these changes stem from a sudden relief of the compressive strain induced by the shell leading to the irreversible relaxation of the Bi core lattice. For a comparison, the electrical properties of the Bi/Te core/shell nanowires are given in the Supplementary Information.

\subsection{Thermoelectric characterization}

\begin{figure}[htbp]
\includegraphics[width=0.95\textwidth]{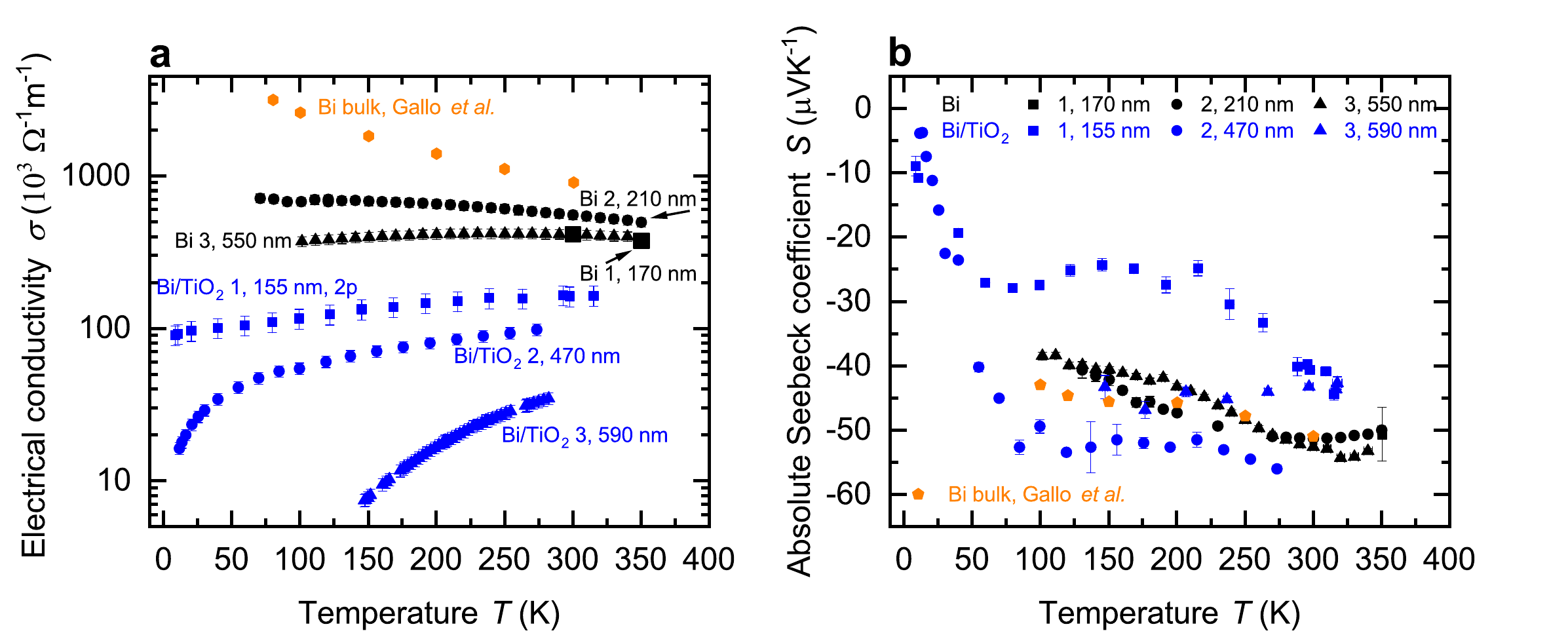}
\caption{\textbf{Electrical conductivity and absolute Seebeck coefficient of the Bi-based core/shell nanowires.} \textbf{a}, Electrical conductivity $\sigma$ of the Bi-based core/shell nanowires as a function of the bath temperature $T$. Bi nanowires exhibit a semimetallic-like electrical conductivity whereas the $\text{Bi}/\text{TiO}_{2}$ nanowires show a semimetallic or seminconducting trend. In addition, the electrical conductivity of Bi bulk (perpendicular to the trigonal axis) from Ref. \cite{Gallo} is added. \textbf{b}, Absolute Seebeck coefficient $S$ of the Bi-based core/shell nanowires as a function of the bath temperature $T$. The absolute Seebeck coefficient of Bi bulk (perpendicular to the trigonal axis) from Ref. \cite{Gallo} is added.}
\label{fig:Bi_Based_Elektrisch_Seebeck}
\end{figure}

The temperature-dependent absolute Seebeck coefficient $S$ of all Bi and $\text{Bi}/\text{TiO}_{2}$ nanowires is shown in Fig. \ref{fig:Bi_Based_Elektrisch_Seebeck}b. $S_{\text{bulk}}$ (perpendicular to the trigonal axis) from Ref. \cite{Gallo} is added to the diagrams. The absolute Seebeck coefficients of all Bi nanowires are comparable with the bulk material in terms of both magnitude and temperature dependence. However, $S$ of $\text{Bi}/\text{TiO}_{2}$ 1 is reduced by $27\,\%$ compared to the bulk material at $T=300\,\text{K}$.

As the absolute Seebeck coefficient $S$ of Bi bulk and of all Bi-based nanowires is negative, electrons are identified as the dominant charge carriers. In general, the total Seebeck coefficient $S_{\text{tot}}$ is determined by the partial contribution of holes and electrons and it is given by  

\begin{equation}
	S_{\text{tot}}=\frac{\sigma_{\text{e}}S_{\text{e}}+\sigma_{\text{h}}S_{\text{h}}}{\sigma_{\text{e}}+\sigma_{\text{h}}}.
	\label{eq:Seebeck_total}
\end{equation}

Here, $\sigma_{\text{e}}$ and $\sigma_{\text{h}}$ are the partial electrical conductivities of the electrons and holes, respectively and $S_{\text{e}}$ and $S_{\text{h}}$ are the partial Seebeck coefficients of the electrons and holes, respectively. Theoretical studies revealed that each partial Seebeck coefficient can be larger than $S_{\text{tot}}$ \cite{Dresselhaus,KimLee} but due to the opposite sign of both contributions, they cancel each other out. This results in a weak temperature dependence between bath temperatures of $T=100\,\text{K}$ and $T=300\,\text{K}$ of $S_{\text{bulk}}$. With decreasing bath temperatures, the absolute Seebeck coefficient is expected to tend to zero.

The temperature dependence and absolute value of $S$ of Bi nanowires are comparable with that of bulk \cite{Gallo}. Theoretical studies showed that Bi nanowires exhibit only a small change of $S$ along the binary axis for diameters between $d=100\,\text{nm}$ and $d=500\,\text{nm}$ \cite{MurataYamamoto}. A significant change of $S$ of Bi nanowires are expected only for diameters below $60\,\text{nm}$ due to a change of the density of states \cite{KimLee}.

For $\text{Bi}/\text{TiO}_{2}$ core/shell nanowires, a contribution of the shell to the total absolute Seebeck can be neglected due to the electrical insulating $\text{TiO}_{2}$. $\text{Bi}/\text{TiO}_{2}$ 1 ($155\,\text{nm}$) has the smallest Seebeck coefficient of all investigated Bi-based nanowires. This can be attributed to the small diameter of the Bi core that is only $95\,\text{nm}$ without the shell. The spatial limitation leads to more surface scattering, a reduced charge carrier mean free path and as a result to a reduction of the absolute Seebeck coefficient \cite{KockertPt,KockertAg}. $\text{Bi}/\text{TiO}_{2}$ 2 ($470\,\text{nm}$) shows an increase of the absolute Seebeck coefficient compared to that of Bi nanowires and Bi bulk. The increase may be attributed to the influence of the compressive strain effect of the $\text{TiO}_{2}$ shell on the Bi core which results in the transition of the semimetallic to semiconducting behavior of the electrical conductivity. For this reason, it is assumed that the Fermi energy is shifted towards the band edge which leads to a higher rate of change in the density of states with energy enhancing the thermoelectric properties. In contrast to $\text{Bi}/\text{TiO}_{2}$ 2 ($470\,\text{nm}$), the Seebeck coefficient of $\text{Bi}/\text{TiO}_{2}$ 3 ($590\,\text{nm}$) is reduced compared to that of the Bi nanowires and of bulk at room temperature. This can be attributed to the band gap opening due to the compressive strain effect of the shell. As a result, the Fermi energy will be shifted towards the band gap beyond the band edge which leads to a lower rate of change in the density of states with energy leading to a reduction of the absolute Seebeck coefficient. Conclusively, the optimum for the Seebeck coefficient for $\text{Bi}/\text{TiO}_{2}$ nanowires with a shell thickness of $30\,\text{nm}$ is a diameter between $400\,\text{nm}$ and $500\,\text{nm}$.

\begin{figure}[htbp]
\includegraphics[width=0.95\textwidth]{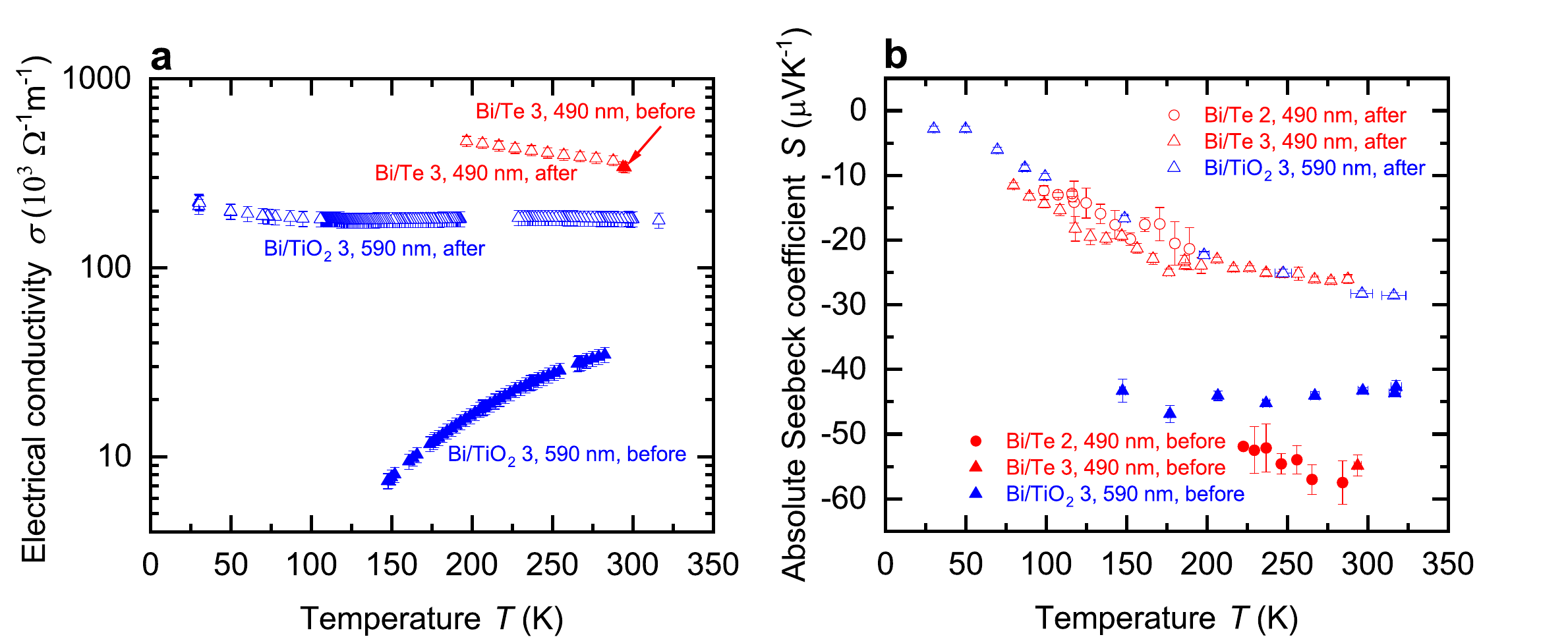}
\caption{\textbf{Influence of lattice relaxation on electrical conductivity and absolute Seebeck coefficient of Bi-based core/shell nanowires.}\textit{Before}, indicates the corresponding transport properties before relaxation. \textit{After}, indicates the corresponding transport properties after relaxation. \textbf{a}, Electrical conductivity $\sigma$ of the Bi-based core/shell nanowires as a function of the bath temperature $T$. The relaxation process of the core/shell structure of $\text{Bi}/\text{TiO}_{2}$ 2 ($470\,\text{nm}$) leads to a change of the temperature dependence of the electrical conductivity from semiconducting to semimetallic. A small change of $\sigma$ can also be observed for Bi/Te 3 ($490\,\text{nm}$). \textbf{b}, Absolute Seebeck coefficient $S$ of the Bi-based core/shell nanowires as a function of the bath temperature $T$. The relaxation process induces a significant change of the absolute Seebeck coefficient of the Bi-based core/shell nanowires. The changes in the transport properties after the relaxation process indicate that the shell had a significant compressive strain effect on the Bi core.}
\label{fig:Bi_Based_vorher_nachher_quer}
\end{figure}

The relaxation process of the Bi core, which changes the electrical conductivity of $\text{Bi}/\text{TiO}_{2}$ 3 ($590\,\text{nm}$), also leads to a significant and irreversible reduction of the Seebeck coefficient, see Fig. \ref{fig:Bi_Based_vorher_nachher_quer}b. As the degree of band energy overlap increases a transition from semiconducting to semimetallic behavior is induced. This is also observed in the Bi/Te nanowires (Supplementary Information).

\subsection{Thermal characterization}

In contrast to the electrical conductivity, all shell materials contribute to the thermal conductivity $\lambda$. Thus, the entire diameter $d$, as given in Tab. \ref{tab:table}, of the nanowires is required to determine $\lambda$. Fig. \ref{fig:Bi_Based_Thermisch_ZT}a shows the thermal conductivity of the Bi and $\text{Bi}/\text{TiO}_{2}$ nanowires and $\lambda_{\text{bulk}}$ (perpendicular to the trigonal axis) from Ref. \cite{Gallo}. The thermal conductivity of all nanowires is reduced compared to the bulk and exhibits a monotonic decrease in $\lambda$ with decreasing bath temperature while Bi bulk shows a monotonic increase. 
In general, $\lambda$ depends on partial contributions from different heat carrier sources and the electronic contribution can be estimated by the Wiedemann-Franz relation \cite{FranzWiedemann,Wilson}. For Bi bulk it has been shown, that phonons are the dominant heat carrier source at low temperatures \cite{Gallo,UherGoldsmid}. As the bath temperatures rises, the charge carrier contribution becomes the dominant part \cite{Gallo,UherGoldsmid}. At $T=300\,\text{K}$, nearly $70\,\%$ of the total thermal conductivity can be attributed to the charge carrier contribution. 
For Bi-based nanowires, the increased surface-area-to-volume ratio acts both on the charge carrier and lattice scattering which leads to a reduction of the thermal conductivity. 

For core/shell nanowires, the shell material has to be taken into account in order to determine the total thermal conductivity $\lambda_{\text{tot}}$ as given by

\begin{equation}
	\lambda_{\text{tot}}=\frac{\lambda_{\text{Bi}}A_{\text{Bi}}+\lambda_{\text{shell}}A_{\text{shell}}}{\lambda_{\text{Bi}}+\lambda_{\text{shell}}}.
	\label{lambda_total}
\end{equation}

$\lambda_{\text{Bi}}$ and $\lambda_{\text{shell}}$ are the partial thermal conductivities of the Bi core and of the shell material, respectively. $A_{\text{Bi}}$ and $A_{\text{shell}}$ are the partial cross-sectional areas of the Bi core and of the shell material, respectively. An upper limit of the thermal conductivity for bismuth oxide is $2.2\,\text{W}\text{m}^{-1}\text{K}^{-1}$ \cite{Anisimova}. $\lambda$ of titanium dioxide films is $1.3\,\text{W}\text{m}^{-1}\text{K}^{-1}$ \cite{AliJuntunen}. The thermal conductivity of the shell yields to a relative uncertainty of the thermal conductivity of the complete nanowires ranging from $<1\,\%$ for Bi 3 ($550\,\text{nm}$) due to large cross-sectional area of the Bi core compared to the small cross-sectional area of the bismuth oxide shell to $14\,\%$ for $\text{Bi}/\text{TiO}_{2}$ 1 ($155\,\text{nm}$) due to the larger cross-sectional area of the $\text{TiO}_{2}$ shell. 
The dominant contribution to the thermal conductivity comes from Bi core. This applies for all core/shell nanowires investigated in this work. 

The thermal conductivity of Bi 2 ($210\,\text{nm}$) is comparable with other Bi nanowires reported in Ref. \cite{KimLee,KimShim}. The reduction of $\lambda$ and the change of the temperature dependence can be attributed to the spatial confinement of the nanowires and the resulting increased phonon surface and charge carrier surface scattering \cite{KimLee,KimShim}. The dominant contribution to the thermal conductivity comes from charge carriers, even when the bath temperatures decreases. Bi 3 ($550\,\text{nm}$) has the smallest $\lambda$ of all Bi-based nanowires investigated in this work.
Scattering at the surface and at grain boundaries reduces the electrical conductivity of Bi 3 ($550\,\text{nm}$) and leads to a strong reduction of the charge carrier contribution to the total thermal conductivity. As a result, the lattice thermal conductivity becomes the main contribution to the total thermal conductivity with decreasing bath temperatures. 

Furthermore, $\lambda$ of Bi 3 ($550\,\text{nm}$) is comparable with that of a Bi/Te core/shell nanowire with a similar diameter but with a rough interface given in Ref. \cite{KangRoh}. The rough surface or interface due to the bismuth oxide may also lead to a reduction of the thermal conductivity. 

The thermal conductivity of $\text{Bi}/\text{TiO}_{2}$ 2 ($470\,\text{nm}$) is reduced compared to the bulk material. The reduction of $\lambda$ can be attributed to an increase of charge carrier and phonon interface scattering comparable to the effect of the Te shell on the thermal conductivity of Bi/Te core/shell nanowires \cite{KangRoh,KimKim,KimOh}. The compressive strain effect of the shell that leads to a reduction of the electrical conductivity will also lead to a reduction of the charge carrier contribution to the total thermal conductivity.

\subsection{Figure of merit}

\begin{figure}[htbp]
\includegraphics[width=0.95\textwidth]{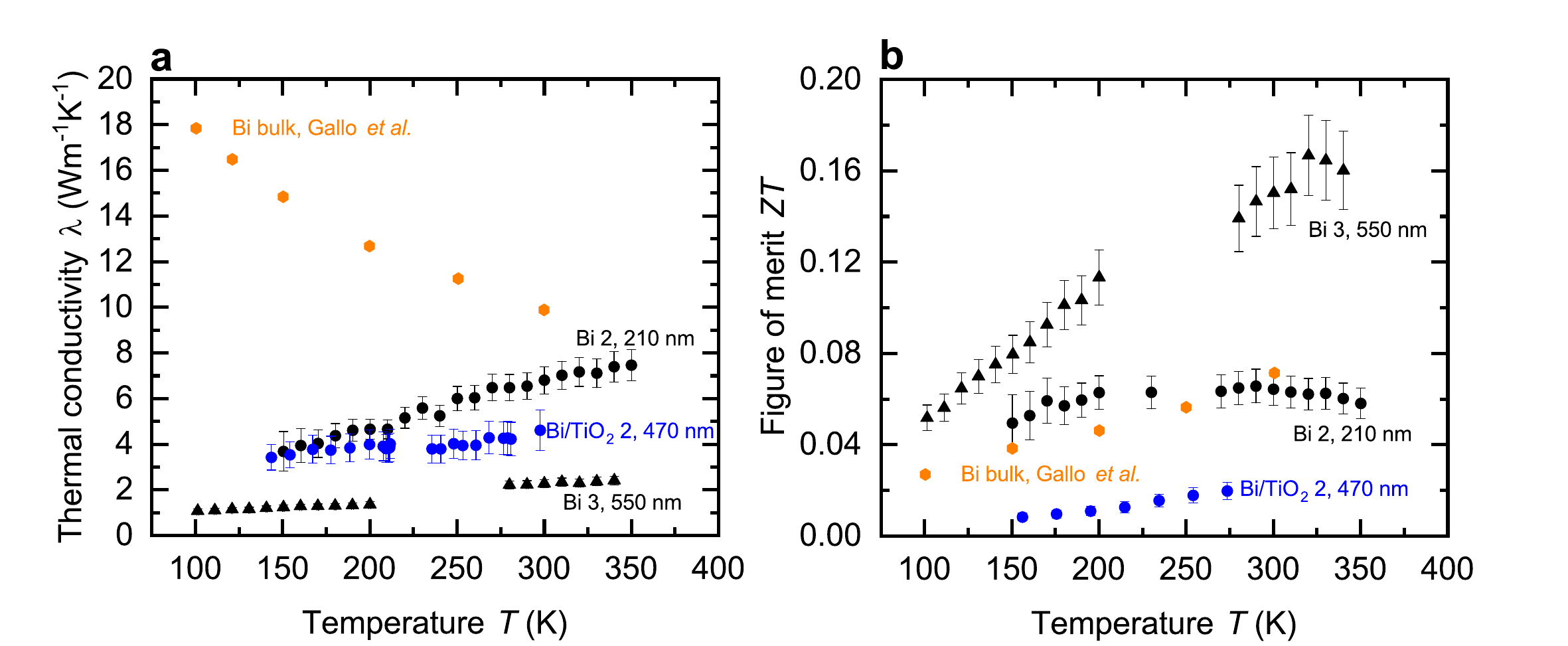}
\caption{\textbf{Thermal conductivity and figure of merit of the Bi-based core/shell nanowires.} \textbf{a}, Thermal conductivity $\lambda$ of the Bi-based core/shell nanowires as a function of the bath temperature $T$. The thermal conductivity of Bi bulk (perpendicular to the trigonal axis) from Ref. \cite{Gallo} is added. $\lambda$ of the Bi-based core/shell nanowires is reduced compared to the bulk material and shows a monotonic decrease of the thermal conductivity with decreasing bath temperature. \textbf{b}, Figure of merit $ZT$ of the Bi-based core/shell nanowires as a function of the bath temperature $T$. The figure of merit of Bi bulk (perpendicular to the trigonal axis) from Ref. \cite{Gallo} is added.}
\label{fig:Bi_Based_Thermisch_ZT}
\end{figure}

The thermoelectric figure of merit $ZT$ at a given bath temperature $T$ is determined by 

\begin{equation}
	ZT=\frac{\sigma S^{2}}{\lambda}T
	\label{figureofmerit}
\end{equation}

and depicted for the Bi-based nanowires in Fig. \ref{fig:Bi_Based_Thermisch_ZT}b. For comparison, the figure of merit of bulk Bi is $ZT=0.07$ at room temperature \cite{Gallo}. This is comparable with $ZT$ of Bi 2 ($210\,\text{nm}$). 

Bi 3 ($550\,\text{nm}$) has the highest figure of merit $ZT=0.15$ at room temperature. This can be attributed to the strong reduction of the thermal conductivity which is probably a result of scattering effects at the surface and in addition at grain boundaries. The thermoelectric properties are tunable over a wide range by the choice of the shell material and the nanowire diameter. For a high figure of merit, it is necessary that the Seebeck coefficient is large enough. The Seebeck coefficient can be increased by shifting the Fermi energy near the band edge. $\text{Bi}/\text{TiO}_{2}$ nanowires show a rather low figure of merit. This can be attributed to the significantly reduced electrical conductivity compared to other Bi-based nanowires due to the influence of the $\text{TiO}_{2}$ shell on the Bi core.

\subsection{Conclusion}

The full temperature-dependent thermoelectric characterization of individual Bi-based core/shell nanowires shows the influence of the shell material on the electrical conductivity, the absolute Seebeck coefficient and the thermal conductivity. Bi-based nanowires are semimetallic or semiconducting depending on the extent of the compressive strain effect induced by the shell. Scattering of charge carriers at surfaces, grain boundaries and core/shell interfaces leads to a reduction of the electrical as well as the thermal conductivity compared to the bulk material. The compressive strain on the Bi core by the shell can increase the Seebeck coefficient by band opening. However, if the strain exceeds the elastic limits, a relaxation process leads irreversibly to a transition from a semiconducting to semimetallic behavior. As a consequence, Bi-based nanowires can be tailored by a shell in a way that the transport properties are tunable over a wide range.

\section*{Conflicts of interest}

There are no conflicts to declare.

\section*{\label{sec:acknowledgments}Acknowledgments}

The authors thank D. Kojda and M. Albrecht for providing the electron beam-induced deposition process for the Bi/Te nanowires. This work emerged from studies within the priority program “Nanostructured Thermoelectrics” SPP 1386 by the German Science Foundation (DFG). We gratefully acknowledge partial funding by DFG, partial financial support by BMBF grant 01DR17012 and HU Berlin. This work was supported by the Agency for Defense Development, Republic of Korea (UD170089GD).

\end{document}


\title[Supplementary Information - Semimetal to semiconductor transition in $\text{Bi}/\text{TiO}_{2}$ core/shell nanowires]{Supplementary Information - Semimetal to semiconductor transition in $\text{Bi}/\text{TiO}_{2}$ core/shell nanowires}

\author{M. Kockert}
\email{kockert@physik.hu-berlin.de}
\affiliation{Novel Materials Group, Humboldt-Universit\"{a}t zu Berlin, 10099 Berlin, Germany}
\author{R. Mitdank}
\affiliation{Novel Materials Group, Humboldt-Universit\"{a}t zu Berlin, 10099 Berlin, Germany}
\author{H. Moon}
\affiliation{Department of Material Science and Engineering, Yonsei University, 03722 Seoul, Republic of Korea}
\author{J. Kim}
\affiliation{Division of Nanotechnology, DGIST, 42988 Daegu, Republic of Korea}
\author{A. Mogilatenko}
\affiliation{Ferdinand-Braun-Institut, Leibniz-Institut f\"{u}r H\"{o}chstfrequenztechnik, 12489 Berlin, Germany}
\author{S. H. Moosavi}
\affiliation{Laboratory for Design of Microsystems, University of Freiburg - IMTEK, 79110 Freiburg, Germany}
\author{M. Kroener}
\affiliation{Laboratory for Design of Microsystems, University of Freiburg - IMTEK, 79110 Freiburg, Germany}
\author{P. Woias}
\affiliation{Laboratory for Design of Microsystems, University of Freiburg - IMTEK, 79110 Freiburg, Germany}
\author{W. Lee}
\affiliation{Department of Material Science and Engineering, Yonsei University, 03722 Seoul, Republic of Korea}
\author{S. F. Fischer}
\email{saskia.fischer@physik.hu-berlin.de}
\affiliation{Novel Materials Group, Humboldt-Universit\"{a}t zu Berlin, 10099 Berlin, Germany}

\maketitle

\section{\label{sec:results}S1: Resistance of Bi-based nanowires}

\begin{figure}[htbp]
\includegraphics[width=0.95\textwidth]{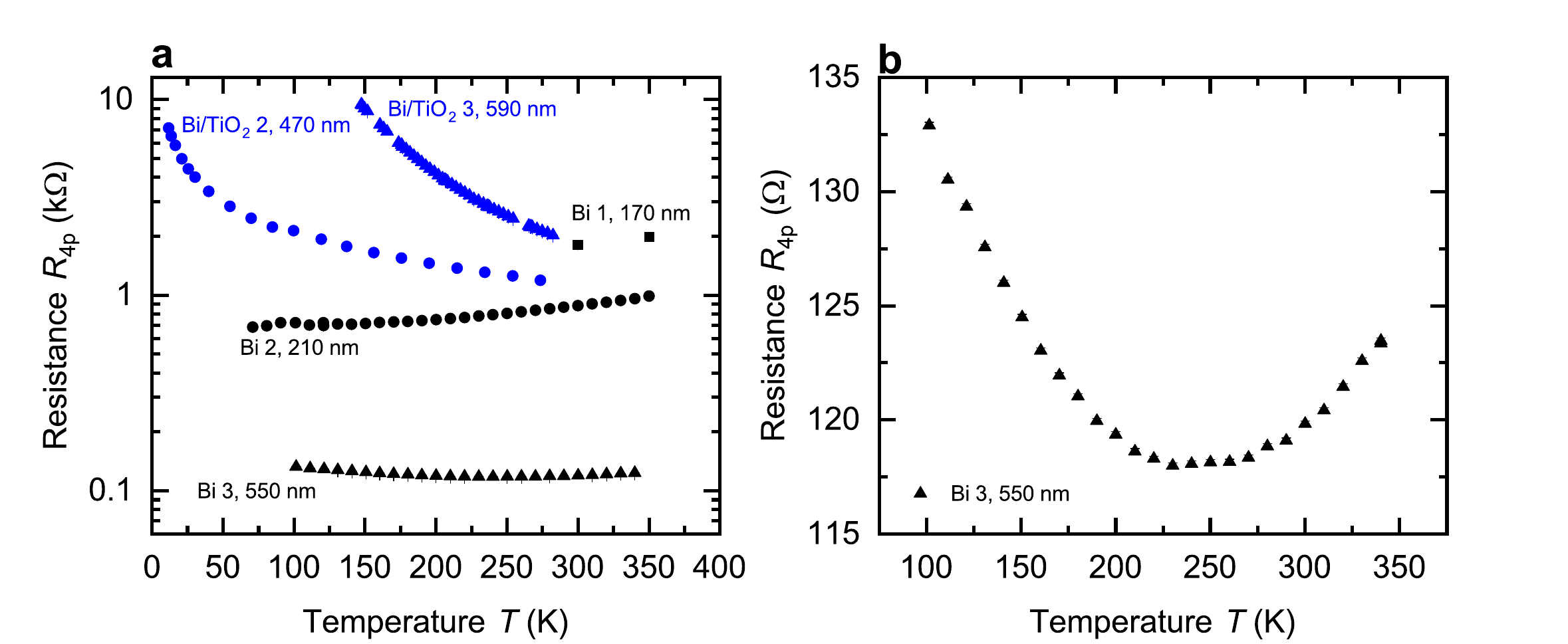}
\caption{\textbf{Resistance of the Bi-based core/shell nanowires.} \textbf{a}, Four-terminal resistance $R_{4\text{p}}$ of the Bi-based core/shell nanowires as a function of the bath temperature $T$. Bi nanowires exhibit a semimetallic temperature dependence. $\text{Bi}/\text{TiO}_{2}$ nanowires show a semiconducting behavior of the resistance. \textbf{b}, Four-terminal resistance $R_{4\text{p}}$ of Bi 3 ($550\,\text{nm}$) as a function of the bath temperature $T$ in a linear representation of $R(T)$. The resistance is decreasing with decreasing bath temperatures from $T=340\,\text{K}$ down to $T=270\,\text{K}$. Below $T=230\,\text{K}$, the resistance is increasing with decreasing temperatures.}
\label{fig:Bi_Based_Widerstand}
\end{figure}

Fig. \ref{fig:Bi_Based_Widerstand}a shows the four-terminal resistance of the Bi-based nanowires as a function of the bath temperature $T$. The Bi nanowires (Bi 1 ($170\,\text{nm}$), Bi 2 ($210\,\text{nm}$) and Bi 3 ($550\,\text{nm}$)) exhibit a semimetallic-like temperature dependence of the resistance. A linear representation of the pronounced temperature dependence of $R$ of Bi 3 is given in Fig. \ref{fig:Bi_Based_Widerstand}b. The $\text{Bi}/\text{TiO}_{2}$ core/shell nanowires show an increase of the resistance with decreasing bath temperatures.

\newpage
\section{\label{sec:results_discussion}S2: Results and discussion}

\subsection{Structural properties}

\begin{figure}[htbp]
\includegraphics[width=0.85\textwidth]{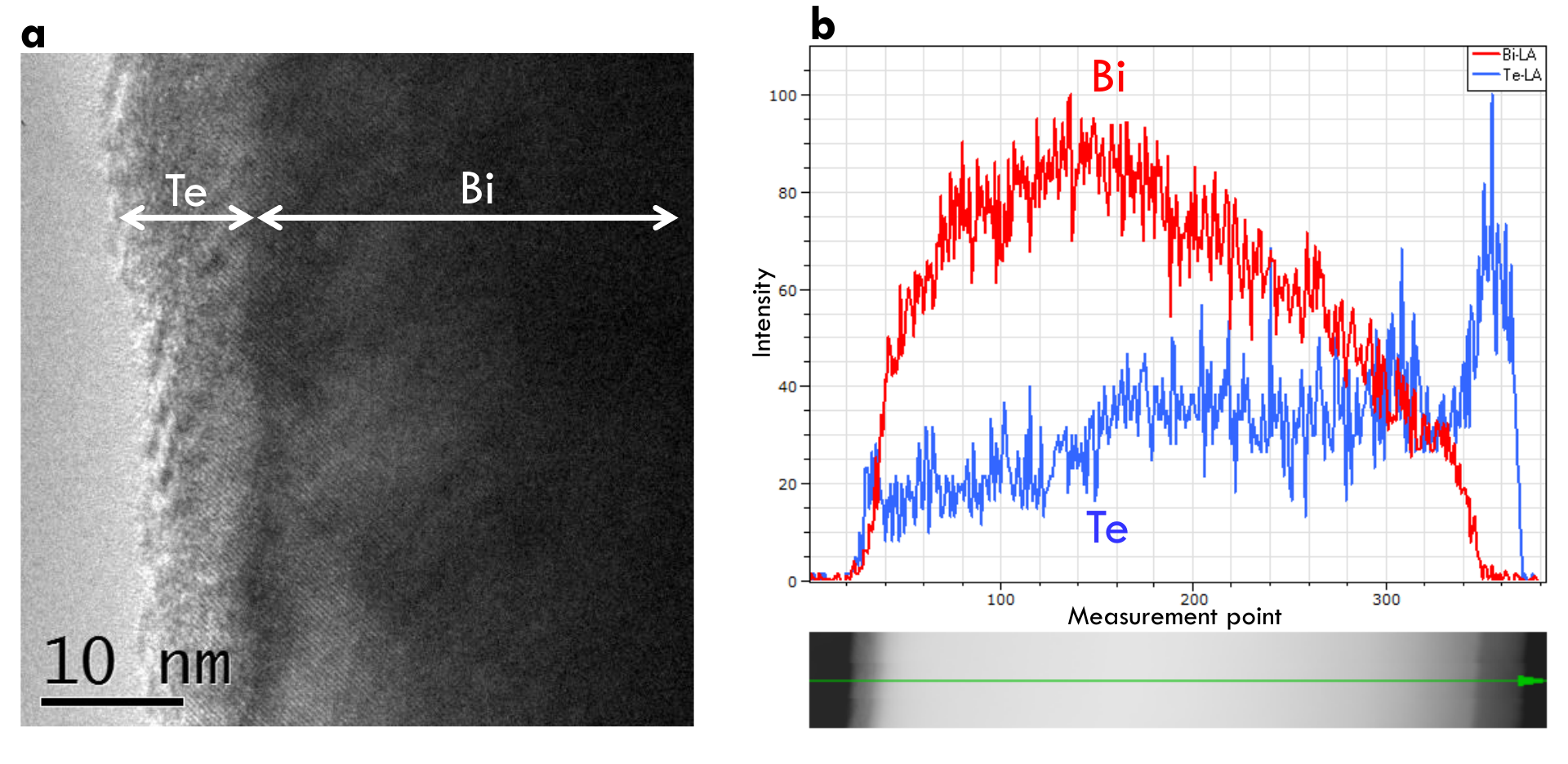}
\caption{\textbf{Structural properties of the Bi/Te core/shell nanowires.} \textbf{a}, Conventional transmission electron microscopy image that shows a section of the Bi/Te nanowire (Bi/Te 1 ($370\,\text{nm}$)). \textbf{b}, Energy dispersive X-ray spectroscopy presenting the tellurium shell distribution of the Bi/Te core/shell nanowire (Bi/Te 1 ($370\,\text{nm}$)). }
\label{fig:REM_TEM_EDX}
\end{figure}

The distinct interface between the tellurium (Te) shell and the bismuth (Bi) core of Bi/Te nanowire (Bi/Te 1 ($370\,\text{nm}$)) can been seen in the conventional transmission electron microscopy image in Fig. \ref{fig:REM_TEM_EDX}a. Elemental line scan obtained across the Bi/Te nanowire by energy dispersive X-ray (EDX) spectroscopy is given in Fig. \ref{fig:REM_TEM_EDX}b. The tellurium shell distribution of the Bi/Te core/shell nanowire (Bi/Te 1 ($370\,\text{nm}$)) is not uniform which indicates the different Te shell thickness on both sides of the nanowire (also see the image obtained by scanning transmission electron microscopy below the EDX line scan). A non-uniform shell can be achieved when the nanowire is not completely perpendicular to the growth substrate during the sputtering process of the shell. In this case the nanowire shadowing effect with respect to the Te adatoms results in formation of a shell with a non-uniform thickness. The influence of the non-uniform shell on the thermoelectric transport properties is discussed later.

\begin{table}[htbp]
	\centering
		\begin{tabular}{|c|c|c|c|}
			\hline 
			Sample & Diameter $d\,(\text{nm})$ & Length $l\,(\mu\text{m})$ & Shell thickness $t\,(\text{nm})$\\
			\hline 
			Bi/Te 1 & $370 \pm 5$ & $21.7 \pm 0.4$ & $10-30$\\
			Bi/Te 2 & $490 \pm 20$ & $12.9 \pm 0.8$ & $10-30$\\
			Bi/Te 3 & $490 \pm 10$ & $12.9 \pm 0.6$ & $10-30$\\
			\hline
		\end{tabular}
		\caption{\textbf{Geometry parameters.} Overview of entire diameter $d$, length $l$ and shell thickness $t$ of bismuth/tellurium (Bi/Te) nanowires. Bi/Te nanowires are coated with a non-uniform Te-shell with a thickness of $10\,\text{nm}-30\,\text{nm}$ by radio frequency sputtering. The geometry parameters have been determined by scanning and transmission electron microscopy.}
		\label{tab:table}
\end{table}

\subsection{Electrical properties}

Fig. \ref{fig:Bi_Te_Transport}a shows the electrical conductivity $\sigma$ of the Bi/Te nanowires as a function of the bath temperature $T$. Moreover, $\sigma_{\text{bulk}}$ (perpendicular to the trigonal axis) from Ref. \cite{Gallo} is added to the diagram. 
Like for Bi nanowire, the Bi/Te core/shell nanowires show also a reduced electrical conductivity compared to the bulk material. Moreover, the shell material can have an additional influence on the electrical conductivity. Kim \textit{et al.} showed that the electrical conductivity of Bi/Te nanowires can be further reduced compared to Bi nanowires and to the bulk material due the compressive strain effect of Te shell on the Bi core \cite{KimKim,KimOh}. This was observed for Bi/Te nanowires with a uniform Te shell thickness. However, the Bi/Te nanowires with a non-uniform shell thickness, which were investigated in this work, showed a reduction of the electrical conductivity compared to the bulk material that is in general not as large as for the Bi/Te nanowires with a uniform Te shell. This can be attributed to the strain effect of the shell on the core that will be larger for a uniform shell thickness than for a non-uniform shell. Furthermore, bismuth and tellurium are both conductive materials. As a result, the electrical conductivity has to be considered as parallel conduction in both materials. The total electrical conductivity of such a combination can be written as

\begin{equation}
	\sigma_{\text{tot}}=\frac{A_{\text{Bi}}\sigma_{\text{Bi}}+A_{\text{Te}}\sigma_{\text{Te}}}{A_{\text{Bi}}+A_{\text{Te}}}.
	\label{eq:sigma_total}
\end{equation}

$A_{\text{Bi}}$ and $A_{\text{Te}}$ are the cross-sectional areas of Bi and Te, respectively. $\sigma_{\text{Bi}}$ and $\sigma_{\text{Te}}$ are the partial conductivities of Bi and Te, respectively. The influence of the Te shell thickness on the reduction of the total electrical conductivity $\sigma_{\text{tot}}$ is illustrated in the following example. Taking Eq. \ref{eq:sigma_total} and the electrical conductivity of Bi bulk $\sigma_{\text{Bi,bulk}}=901600\,\Omega^{-1}\text{m}^{-1}$ \cite{Gallo} and Te bulk $\sigma_{\text{Te,bulk}}=185\,\Omega^{-1}\text{m}^{-1}$ \cite{Nussbaum} and assuming the cross-sectional area of a nanowire with a total diameter of $300\,\text{nm}$ results in a reduction of $\sigma_{\text{tot}}$ by $13\%$ compared to $\sigma_{\text{Bi,bulk}}$ if the Te shell thickness is $10\,\text{nm}$ or in a reduction of $\sigma_{\text{tot}}$ by $36\%$ if the Te shell thickness is $30\,\text{nm}$. As a result, a uniform Te shell will lead to a larger reduction of $\sigma$ compared to the bulk than a non-uniform shell.

\subsection{Thermoelectric properties}

Fig. \ref{fig:Bi_Te_Transport}b shows the absolute Seebeck coefficient $S$ of the Bi/Te nanowires as a function of the bath temperature $T$. $S_{\text{bulk}}$ (perpendicular to the trigonal axis) from Ref. \cite{Gallo} is added to the diagram. In general, for Bi/Te core/shell nanowires the direct influence of the Te shell on the total Seebeck coefficient as part of a parallel conduction model can be neglected due to the larger cross-sectional area and electrical conductivity of the Bi core compared to the Te shell. $S$ of the Bi/Te core/shell nanowires is larger than that of Bi bulk. The absolute Seebeck coefficient of Bi/Te 1 ($370\,\text{nm}$) is increased by $27\,\%$ compared to the bulk material at $T=290\,\text{K}$ and it 
has the largest $S$ of all investigated Bi-based nanowires in this work. This can be attributed to the compressive strain effect of the Te shell on the Bi core as previously reported in Ref. \cite{KimKim,KimOh}. However, $S$ is smaller compared to the data given in Ref. \cite{KimKim,KimOh}. This can be explained by the non-uniform shell of the Bi/Te nanowires, as shown in Fig. \ref{fig:REM_TEM_EDX}a,b, and the resulting lower compressive strain effect of the Te shell on the Bi core compared to Bi/Te nanowires with a uniform shell. 

\subsection{Thermal properties}

\begin{figure}[htbp]
\includegraphics[width=0.55\textwidth]{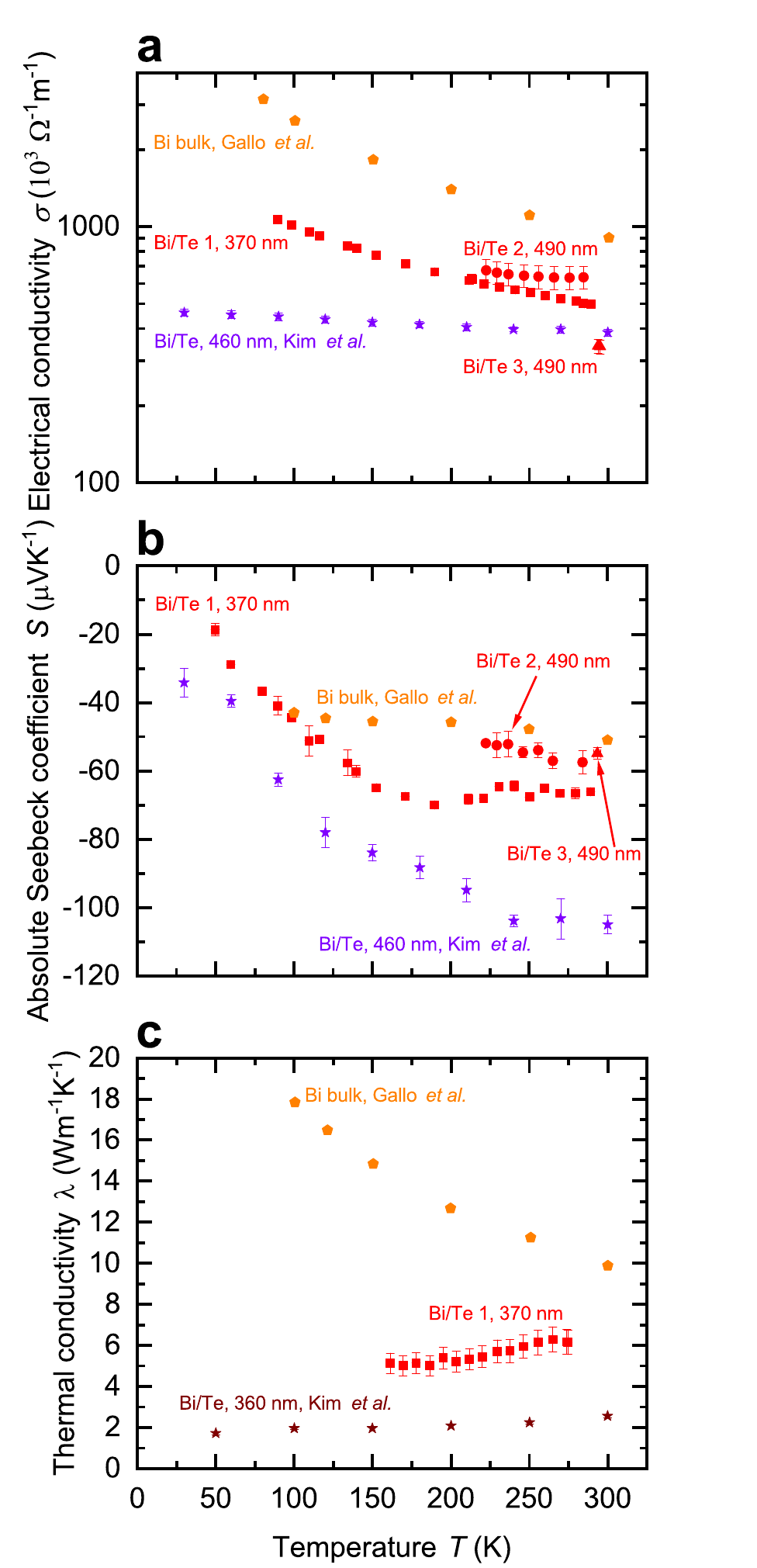}
\caption{\textbf{Electrical conductivity, absolute Seebeck coefficient and thermal conductivity of the Bi/Te core/shell nanowires with non-uniform Te shell.} \textbf{a}, Electrical conductivity $\sigma$ of the Bi/Te core/shell nanowires as a function of the bath temperature $T$. $\sigma$ is reduced compared to the bulk material but increased compared to a Bi/Te nanowire with a uniform Te shell. \textbf{b}, Absolute Seebeck coefficient $S$ of the Bi/Te core/shell nanowires as a function of the bath temperature $T$. The modulus of $S$ is increased compared to the bulk material but decreased compared to a Bi/Te nanowire with a uniform Te shell. \textbf{c}, Thermal conductivity $\lambda$ of the Bi/Te core/shell nanowires as a function of the bath temperature $T$. The thermal conductivity of the Bi/Te nanowires is reduced compared to the bulk material and exhibits an opposite temperature dependence. The transport properties of Bi bulk (perpendicular to the trigonal axis) from Ref. \cite{Gallo} and of different Bi/Te nanowires with a uniform shell from Ref. \cite{KimOh} are added to the corresponding diagram. The Seebeck coefficient of the of Bi/Te nanowire from Ref. \cite{KimOh} was corrected by the absolute Seebeck coefficient of the reference material.}
\label{fig:Bi_Te_Transport}
\end{figure}

For the Bi/Te nanowires it was shown, that the rough interface between the Bi core and Te shell can lead to a reduction of the thermal conductivity as reported in Ref. \cite{KangRoh,KimKim,KimOh}. However, the thermal conductivity of Bi/Te 1 ($370\,\text{nm}$) is larger compared to other Bi/Te core/shell nanowires reported in Ref. \cite{KangRoh,KimKim,KimOh}. This can be attributed to the non-uniform Te shell as shown in Fig. \ref{fig:REM_TEM_EDX}a,b. As a result, the compressive strain effect is lower compared to Bi/Te nanowires with a uniform shell. This will lead to a larger charge carrier contribution to the thermal conductivity increasing the overall thermal conductivity.
Fig. \ref{fig:Bi_Te_Transport}c shows the thermal conductivity $\lambda$ of the Bi/Te nanowires as a function of the bath temperature $T$. Moreover, $\lambda_{\text{bulk}}$ (perpendicular to the trigonal axis) from Ref. \cite{Gallo} is added to the diagram. 



